**Complex nonlinear capacitance in outer hair cell macro-patches: effects of membrane tension.**

Joseph Santos-Sacchi [1, 2, 3] and Winston Tan [1]

[1] Surgery (Otolaryngology), [2] Neuroscience, and [3] Cellular and Molecular Physiology, Yale University School of Medicine, 333 Cedar Street, New Haven, CT 06510, USA

*Running title:* tension effects of OHC complex NLC frequency response

*Send correspondence to:*

        Joseph Santos-Sacchi
        Surgery (Otolaryngology), Neuroscience, and Cellular and Molecular Physiology
        Yale University School of Medicine
        BML 224, 333 Cedar Street
        New Haven, CT  06510
        Phone:  (203) 785-7566
        e-mail:  joseph.santos-sacchi@yale.edu

**Acknowledgments:**

This research was supported by NIH-NIDCD R01 DC000273, R01 DC016318 and R01 DC008130 to J.S.-S.

Abstract

Outer hair cell (OHC) nonlinear capacitance (NLC) represents voltage sensor charge movements of prestin (SLC26a5), the protein responsible for OHC electromotility. Previous measures of NLC frequency response have employed methods which did not assess the influence of dielectric loss (sensor charge movements out of phase with voltage) that may occur, and such loss conceivably may influence prestin's frequency dependent activity. Here we evaluate prestin's complex capacitance out to 30 kHz and find that prestin's frequency response determined using this approach coincides with all previous estimates. We also show that membrane tension has no effect on prestin's frequency response, despite substantial shifts in its voltage operating range, indicating that prestin transition rate alterations do not account for the shifts. The magnitude roll-off of prestin activity across frequency surpasses the reductions of NLC caused by salicylate treatments that are known to abolish cochlear amplification. Such roll-off must therefore limit the effectiveness if prestin in contributing to cochlear amplification at the very high acoustic frequencies processed by some mammals.

Prestin (SLC26a5) underlies outer hair cell (OHC) mechanical activity (Zheng et al., 2000), whereby voltage-dependent conformational transitions couple into length changes of the cell (electromotility; eM) (Kachar et al., 1986; Santos-Sacchi and Dilger, 1988; Liberman et al., 2002). Sensor charge movements associated with these conformational changes are measurable as an electrical correlate of eM, i.e., nonlinear capacitance (NLC) (Ashmore, 1990; Santos-Sacchi, 1990, 1991), which is maximal at $V_h$, the voltage where prestin charge is distributed equally across the OHC membrane and where eM gain is greatest.

The study of OHC NLC by admittance techniques in whole cell voltage clamp is compromised by contributions from stray capacitance, membrane conductances, and electrode series resistance ($R_s$), the latter altering over time due to plugging of the patch pipette tip with intracellular constituents. Because of this, it is virtually impossible to measure complex membrane capacitance in whole cell mode. We avoid these issues by measuring membrane admittance from macro-patches of the OHC lateral membrane. The lateral membrane of the cylindrical OHC is dominantly populated by prestin (> 8000 functional units/ $\mu m^2$ (Huang and Santos-Sacchi, 1993; Gale and Ashmore, 1997b; Mahendrasingam et al., 2010)), whereas voltage-dependent membrane conductances are housed in the basal pole membrane (Santos-Sacchi et al., 1997). Consequently, macro-patch admittance, following removal of stray capacitance by subtraction of admittance at very depolarized levels where NLC is absent (Santos-Sacchi and Navarrete, 2002), can be used to study complex sensor-charge movements arising from voltage-induced conformational changes in prestin.

Here, we evaluate patch admittance to provide estimates of complex capacitance representing charge movements both in phase and 90° out of phase with AC voltage excitation. We compare such data to previously obtained measures of OHC NLC frequency response (Gale and Ashmore, 1997a; Santos-Sacchi and Tan, 2019), and additionally report on the effects of membrane tension (Iwasa, 1993; Gale and Ashmore, 1994; Kakehata and Santos-Sacchi, 1995) on the frequency response of complex NLC. Will an assessment of NLC based on measures of complex NLC alter our current view, as suggested (Sun et al., 2009), and can turgor pressure, normally present in the OHC (Ratnanather et al., 1993), with its effect on membrane tension be influential? Our observations indicate that the bulk of sensor charge movement is in phase with voltage, while the resistive component (dielectric loss) is relatively small. Furthermore, membrane tension, though altering prestin's operating voltage point, has no effect on its frequency response. Thus, magnitude estimates of complex NLC are comparable to those measured with other methods (Gale and Ashmore, 1997a; Santos-Sacchi and Tan, 2019), being unusually low pass in nature (non-Lorentzian) and indicating that the absolute movement of prestin charge that drives electromotility (eM) (Wu and Santos-Sacchi, 1998; Santos-Sacchi and Tan, 2018) is unlikely to extend with high fidelity to the very high acoustic frequencies (60-160 kHz) detected by some mammals.

**Methods**

Methods, including details of our voltage chirp stimulus protocol, are detailed in (Santos-Sacchi and Tan, 2019). Briefly, extracellular solution was (in mM): NaCl 100, TEA-Cl 20, CsCl 20, $CoCl_2$ 2, $MgCl_2$ 1, $CaCl_2$ 1, Hepes 10, pH 7.2. Extracellular solution was in the patch pipette. Macro-patches on the OHC lateral membrane were made near the middle of the cylindrical cell, since prestin density/activity is uniform within the lateral membrane (Dallos et al., 1991; Huang and Santos-Sacchi, 1993). Borosilicate pipettes of inner tip dimeters between 3-4 μm were used, with M-coat applied within about 20 μm of the tip to minimize pipette capacitance. In order to establish gigohm seals we supplemented extracellular solution with 5-7.5 μM $Gd^{+3}$; we have shown previously that theses low concentrations help to form seals without affecting NLC (Santos-Sacchi and Song, 2016; Santos-Sacchi et al., 2019a). An Axon 200B amplifier was used with jClamp software (www.scisoftco.com). An Axon Digidata 1440 was used for digitizing at 10 μs (Nyquist frequency of 50 kHz). Membrane admittance was determined using a series of voltage chirps (4096 points each, resolution 24.4 Hz) superimposed onto holding potentials ranging from -160 to 160 mV, in 40 mV increments. 100 ms after each step to a new holding potential, 26 contiguous chirp-induced current responses at each holding potential were time-averaged. Real and imaginary components of the membrane admittance at all chirp frequencies were determined by FFT in jClamp, and corrected for the roll-off of recording system admittance (Gillis, 1995; Santos-Sacchi and Tan, 2019). Our dual-sine methodology permitted us to measure capacitance with high fidelity out to 20 kHz across holding potentials (Santos-Sacchi and Tan, 2019). Here using single sine analysis we extend that to 30 kHz.

Our patch pipette inner tip diameters (see **Results**) were used to estimate the linear capacitance of membrane patches. We estimate a linear membrane patch capacitance of 187.8 +/- 15.4 fF (n=25). This was determined by estimating membrane patch hemispheric surface area using the standard value of 1 μF/ $cm^2$ (Hille, 1992). In our presentation below, we provide absolute estimates of NLC, and specific estimates of NLC by dividing patch admittance with linear capacitance for each patch, thereby accounting for different patch size. Patch experiment data were accepted for inclusion if maximum NLC within our recording bandwidth was > 150 fF. Complex values were first averaged for all average-based analyses. Plot traces are smoothed with a 6 point (150 Hz bandwidth) running average in Matlab.

In order to extract Boltzmann parameters, capacitance-voltage data were fit to the first derivative of a two-state Boltzmann function.

$$C_m = NLC + C_{sa} + C_{\text{lin}} = Q_{\max} \frac{ze}{k_B T} \frac{b}{(1+b)^2} + C_{sa} + C_{\text{lin}} \qquad (m1)$$

$$\text{where} \quad b = exp\left(-ze\frac{V_m - V_h}{k_B T}\right), \ C_{sa} = \frac{\Delta C_{sa}}{(1+b^{-1})}$$

$Q_{max}$ is the maximum nonlinear charge moved, $V_h$ is voltage at peak capacitance or equivalently, at half-maximum charge transfer, $V_m$ is $R_s$-corrected membrane potential, $z$ is valence, $C_{\text{lin}}$ is linear membrane capacitance, e is electron charge, $k_B$ is Boltzmann's constant, and T is absolute temperature. $C_{sa}$ is a component of capacitance that characterizes sigmoidal changes in specific membrane capacitance (Santos-Sacchi and Navarrete, 2002; Santos-Sacchi and Song, 2014a). Functional prestin density in the membrane is based on quantity of sensor charge.

$$Q_{max} = \left[\frac{4 \cdot k_B \cdot T}{z \cdot e}\right] \cdot NLC_{Vh} \qquad (m2)$$

$$Prestin\ density = \frac{Q_{max}}{patch\ surface\ area} \qquad (m3)$$

A power fit of NLC across frequency ($f$) was performed.

$$NLC(f) = NLC_0 + a * f^b \qquad (m4)$$

where $NLC_0$ is the zero frequency component, and $a$ and $b$ control the frequency response.

**Results**

In our present study, we successfully recorded from 25 macro-patches on the OHC lateral membrane, where membrane breakdown did not occur; breakdown might be expected beyond our voltage protocol voltages (Navarrete and Santos-Sacchi, 2006). Our pipette inner tip diameter was 3.45 +/- 0.15 $\mu$m. Given a typical resistivity for pipette solutions of 100 ohm-cm (Mathias et al., 1990), and our pipette taper angle of 0.2 rad, we calculate an $R_s$ of about 920 k$\Omega$. Since the patch membrane extends into the pipette where pipette diameter is larger, we further estimate a reduction of $R_s$ to 735 k$\Omega$.

The lateral membrane of the OHC is virtually devoid of voltage-dependent conductances (Santos-Sacchi et al., 1997), and we additionally use channel blockers to insure this. Our seal resistance, determined by linear fit of step induced currents within a linear region of the I-V between -40 and + 40 mV, was 5.39 +/- 0.65 G$\Omega$.

Consequently, the macro-patch membrane, unlike the membrane in whole cell conditions, may be considered an isolated capacitor under voltage clamp, and thus amenable to determination of complex capacitance. In the following analysis of complex capacitance, we follow the methodology of Fernandez et al. (Fernandez et al., 1982a), applied to each patch individually, based on its characteristics. The OHC patch capacitance presents as a parallel combination of linear ($C_{lin}$) and prestin-generated (NLC) capacitance. Conceivably, prestin's voltage-sensor may work as an imperfect, lossy capacitor that possesses both resistive (due to dielectric loss) and capacitive components (e.g., modelled as a combination of capacitor and resistor); that imperfection may influence estimates of NLC (Sun et al., 2009). In addition to biophysical capacitance, system-generated stray capacitance ($C_{stray}$) will contribute to our measures. $C_{stray}$, though voltage-independent, may also possess resistive and capacitive components, its admittance being $Y^*_{stray}(\omega)$. Under voltage clamp, an AC voltage across the patch membrane ($V_m$) induces an AC current ($I_m$), where the admittance ($Y_m = I_m/V_m$) is a complex function of angular frequency, $\omega = 2\pi f$ and $i = \sqrt{-1}$

$$Y_m(\omega) = G_m(\omega) + iB_m(\omega) \qquad (1)$$

with $G_m$ representing membrane conductance, $B_m$ representing membrane suseptance. Before continuing, we remove the effects of series resistance ($R_s$) by subtracting it from the real component of membrane impedance $Z_m(\omega)$ ($1/Y_m(\omega)$), and then converting back to admittance (Fernandez et al., 1982a). $Y_m(\omega)$ can be described in more detail,

$$Y_m(\omega) = G_m(\omega) + G_{leak} + Y^*_{stray}(\omega) + i\omega\, C_m(\omega) \qquad (2)$$

$$\text{where } C_m(\omega) = C_{lin} + NLC.$$

$G_{leak}$ represents a DC leakage conductance. $C_{lin}$ is taken as frequency independent and via small signal analysis we seek to determine the frequency dependence of NLC, after removing $Y^*_{stray}(\omega)$. To our benefit, admittance at + 160 mV lacks NLC (Santos-Sacchi and Navarrete, 2002)

$$Y^{160}_m(\omega) = G_m(\omega) + G_{leak} + Y^*_{stray}(\omega) + i\omega\, C_{lin} \qquad (3)$$

Thus, subtraction of membrane admittance at +160 mV from those corresponding measures at all other holding potentials provides a differential admittance, $Y^*_m(\omega)$, devoid of stray capacitance effects.

$$Y^*_m(\omega) = Y_m(\omega) - Y^{160}_m(\omega) = G^*_m(\omega) + iB^*_m(\omega) \qquad (4)$$

Actually, after subtraction, there remains a small differential residual nonlinear, voltage-dependent DC leakage conductance ($G_{leak}$), which we remove from $G^*_m(\omega)$ by subtraction of the real part of $Y^*_m(\omega)$ at zero frequency. We determine this value to subtract by extrapolating to zero frequency with a linear fit of Re($Y^*_m(\omega)$) between 24.41 and 463.86 Hz at each of the stepped holding potentials. This is akin to removing leakage conductance determined by prior DC step estimates, but has the further advantage of being determined during the actual chirp stimulation period.

Complex membrane capacitance, a function of angular frequency, is defined as (see Fernandez et al. (Fernandez et al., 1982a) )

$$C^*_m(\omega) = \frac{Y^*_m(\omega)}{i\omega} = \frac{B^*_m(\omega)}{\omega} + i\,\frac{G^*_m(\omega)}{\omega} = C^{*\prime}_m(\omega) - i\, C^{*\prime\prime}_m(\omega) \qquad (5)$$

In **Figure 1A**, we plot the average capacitive (real, $C^{*\prime}_m(\omega)$) and apparent conductance (imaginary, $C^{*\prime\prime}_m(\omega)$) components of the complex capacitance for holding potentials of 40, 0, -40 and -80 mV. The values are voltage dependent. While the capacitive component is large and frequency dependent, the conductance component is smaller and less frequency dependent, differing from a larger component predicted from diffusion-based charge translocation (Sun et al., 2009).

The absolute magnitude of the complex capacitance can be used to glean an estimate of membrane capacitance ($C_m^1(\omega)$),

$$C_m^1(\omega) = |C_m^*(\omega)| \qquad (6)$$

In **Figure 1B**, the absolute magnitude, $|C_m^*(\omega)|$, of the complex capacitance is plotted. The magnitude rolls off continuously across frequency at each holding potential. In **Figure 1C**, the data are plotted as a specific complex capacitance, i.e., per linear capacitance of the patches. As found in whole cell recordings, peak NLC can be larger than linear capacitance (Ashmore, 1990; Santos-Sacchi, 1991).

It is also possible to estimate capacitance of the macro-patch membrane by a "phase tracking" approach. That is, the phase angle of the complex admittance $Y_m^*(\omega)$ can be rotated to minimize its real component ($G_m^*(\omega)$), providing a new value ($Y_m^{**}$), whose imaginary component reflects the capacitive component (suseptance) of the admittance in the absence of conductance interference. Another estimate of membrane capacitance, $C_m^2(\omega)$, can then be obtained.

$$\angle\left(Y_m^*(\omega)\right) = \operatorname{atan}\left(\frac{B_m^*(\omega)}{G_m^*(\omega)}\right) \qquad (7)$$

$$Y_m^{**} = Y_m^*(\omega) \cdot e^{\left(-i\left(\angle\left(Y_m^*(\omega)\right)-\frac{\pi}{2}\right)\right)} \qquad (8)$$

$$C_m^2(\omega) = \frac{\operatorname{Im}(Y_m^{**})}{\omega} \qquad (9)$$

This approach is similar to traditional real time capacitance phase tracking under voltage clamp (Fidler and Fernandez, 1989), where the capacitive component at the angle of $\delta Y/\delta C_m$ (Santos-Sacchi, 2004) is obtained by adjusting the lock-in recording angle until the conductance component is minimized and the capacitive component is maximized. In that approach, calibration with a known capacitance provides membrane capacitance estimates at the measurement frequency. This approach to measure OHC patch NLC was used by Gale and Ashmore (Gale and Ashmore, 1997a).

In **Figure 2 A**, we plot the two estimates of OHC NLC [$C_m^1(\omega)$, and $C_m^2(\omega)$], corresponding to methods utilizing the complex capacitance magnitude, and the phase tracking approach, respectively – both at -40 mV holding potential (near $V_h$). The measures overlap, indicating that *eqs. 6* and *9* return the same result. In **Figure 2 B**, the voltage dependence of the complex capacitance magnitude at selected frequencies, presents a bell-shaped function typical of OHC NLC (Ashmore, 1990; Santos-Sacchi, 1991), whose peak precipitously decreases with frequency, but whose voltage at peak ($V_h$) remains similar across frequency. These data are fit (*eq. m1*) to extract the Boltzmann parameters given in the legend. Prestin density based on fits to capacitance at 1 kHz is 1133/$\mu m^2$, similar to previous estimates at a similar frequency (Huang and Santos-Sacchi, 1993; Gale and Ashmore, 1997b; Mahendrasingam et al., 2010).

Having arrived at practical approaches to estimate OHC patch NLC, we now look at the influence of membrane tension. Several studies under whole cell voltage clamp have found that as membrane tension is increased, NLC $V_h$ shifts in the depolarizing direction, and decreases in peak magnitude (Iwasa, 1993; Gale and Ashmore, 1994; Kakehata and Santos-Sacchi, 1995). However, utilizing lock-in estimates of lateral membrane patch capacitance instead of the whole cell technique, Gale and Ashmore (Gale and Ashmore, 1997b) found no evidence for a reduction in peak capacitance despite shifts in the depolarizing direction.

In **Figure 3**, we explore the effects of membrane tension on the magnitude, $|C_m^*(\omega)|$, of complex NLC across frequency, as it is a robust measure of NLC, equivalent to that obtained with the phase tracking approach. Results at 0, -4, -8 and -10 mm Hg (i.e., 0, 0.53, 1.06 and 1.33 kPa) pipette pressure are shown in **Figs. A-D,** left panels. Membrane tension shifts $V_h$, as indicated by the rearrangement of the capacitance magnitude traces as negative pressure alters. For example, in **left panel A**, the trace at -80 mV (green) is above the trace at 0 mV (red), whereas in **left panel D**, the positions are reversed. In the right panels, plots of C-V functions of complex capacitance magnitude at the four different pipette pressures are shown (mean +/- SEM, n=8). The depolarizing shift in $V_h$ is readily apparent as negative pressure increases (red bars). There is little change in peak capacitance, however. $\Delta C_{sa}$ is mainly unaffected by tension.

For comparison of frequency dependence as a function of membrane tension, **Figure 4 A** replots the peak magnitude traces (-40 mV) from **Figure 3** at the different pipette pressure. The frequency response of complex capacitance magnitude is unaffected by membrane tension, indicated by the overlap of traces. The sensitivity of patch $V_h$ shift to tension was 24.1 mV/kPa based on our fits (**Figure 4 B**).

We previously measured OHC NLC in macro-patches using a dual-sinusoidal methodology (Santos-Sacchi and Tan, 2019). In **Figure 5**, we replot those $NLC_{vh}$ data (blue line) and a power function of frequency fit of those data (red line), alongside $NLC_{vh}$ data from Gale and Ashmore (Gale and Ashmore, 1997a). Their data (green circles, SEM, and fit from their Fig. 3B; in green) are shown scaled (X 3.5) to overlie our data. Their measures were made with a lock-in amplifier using a 100 fF calibration at each frequency. Scaling to our data was necessary because our average patch size/capacitance was greater than theirs. As is evident, our data fit intersects their data points, indicating an acceptable correspondence to the extended power fit frequency response of our data. Finally, extraction of NLC$_{vh}$ from our new measures (red circles) and power fit (magenta line) of patch absolute complex capacitance (from **Figure 3**, scaled X 1.05) corresponds well to our previous observations and theirs. We note that the absolute magnitude of complex capacitance $(C_m^1(\omega))$ is equivalent to our phase tracking approach $(C_m^2(\omega))$ to estimate capacitance (**Figure 2**), each corresponding to the estimates provided by the traditional phase tracking method (Neher and Marty, 1982) employed by Gale and Ashmore (Gale and Ashmore, 1997a). Furthermore, since each method takes into account the dielectric loss in prestin, we conclude that this dielectric loss does not significantly account for the low-pass nature of prestin charge movement. To further substantiate this claim, we make another estimate of membrane capacitance, $C_m^3(\omega)$, one that does not take into account dielectric loss, but is appropriate for a loss-less capacitor.

$$C_m^3(\omega) = \frac{Im(Y_m^*(\omega))}{\omega} = \frac{B_m^*(\omega)}{\omega} \tag{10}$$

**Figure 6** shows the results for our three estimates of prestin capacitance [$C_m^1(\omega)$, $C_m^2(\omega)$ and $C_m^3(\omega)$], all reasonably overlying each other, and exhibiting low-pass behavior. Thus, all results available to date show an unusually low-pass behavior of prestin voltage-sensor performance, with little influence of dielectric loss.

**Discussion**

We previously evaluated OHC macro-patch NLC frequency dependence utilizing a dual-sine methodology that worked on extracted prestin displacement currents (Santos-Sacchi and Tan, 2019). That approach did not allow separation of real and imaginary components of NLC. Here we provide an analysis of OHC NLC in membrane macro-patches utilizing single sine methods that have been used to characterize gating charge movements in other voltage-dependent proteins (Fernandez et al., 1982a), where real (capacitive) and imaginary ("resistive", representing dielectric loss) components of sensor charge are separated. We explored these alternative approaches in order to determine whether new estimates of NLC frequency response incorporating the influence of dielectric loss would substantially alter our previous estimates, as was predicted through modelling (Sun et al., 2009). This is not the case, however. Thus, the bulk of prestin charge movement appears to move in phase with voltage, and the small dielectric loss that we find does not significantly influence measures of NLC frequency response.

It was previously noted that low frequency determined OHC NLC (C-V) measures are unable to identify whether prestin charge movement is multistate/diffusional in nature (Huang and Santos-Sacchi, 1993; Scherer and Gummer, 2005). However, wide-band frequency interrogation of prestin NLC may provide clues. In this regard, the large bell-shaped (across frequency) resistive component of complex capacitance found for DPA$^-$ which moves through bulk membrane lipid (Lu et al., 1995), and which Sun et al. (Sun et al., 2009) predicted for prestin modelled as a simple diffusion process of charge translocation is not found here. The resistive component we measure is relatively small and flat across frequency (**Figure 1A**, lower panel). This is likely because prestin charge movement is not a diffusional process, but similar to gating charge movement in other membrane proteins. Thus, Fernandez et al. (Fernandez et al., 1982b) found that whereas chloroform, which alters membrane viscosity, can influence the speed of DPA$^-$ charge translocation within the membrane, Na channel voltage-sensor charge translocation speed is not affected. Similarly, sensor charge in prestin is likely constrained to charged protein residues that move during protein conformational change (Bai et al., 2009). Indeed, Sun et. al.'s model included membrane diffusion of chloride ions as voltage sensors in prestin. However, chloride anions do not solely function as extrinsic voltage sensors for prestin, nor do they move through bulk lipid; instead they likely influence prestin kinetics (Rybalchenko and Santos-Sacchi, 2003; Song and Santos-Sacchi, 2010; Santos-Sacchi and Tan, 2019) (also see our comment #1 on Walter et. al eLife 8, 2019 - https://elifesciences.org/articles/46986) (Walter et al., 2019).

Prestin's voltage operating range, whose midpoint is at $V_h$, is very sensitive to membrane tension (Iwasa, 1993). Here, we measured the effect of membrane tension on prestin performance by changing patch pipette pressure to study its influence on NLC frequency response. The average sensitivity of patch $V_h$ shift to tension was 24.1 mV/kPa, which corresponds well to our previous measures (Kakehata and Santos-Sacchi, 1995) in whole cell mode of 21.6 mV/kPa. Adachi et al. (Adachi et al., 2000) found similar sensitivity in whole cell mode (27.5 mV/kPa). Gale and Ashmore (Gale and Ashmore, 1997b), in their patches, found a somewhat smaller sensitivity of 11.1 mV/kPa. In prestin transfected cells, values of near 4 mV/kPa were found (Ludwig et al., 2001; Santos-Sacchi et al., 2001). One mechanism whereby shifts in $V_h$ might occur would be due to changes in the ratio of transition rates between its conformational states. Naturally, if this were occurring then changes in NLC frequency response might arise. We found no change in NLC magnitude or frequency response during alterations in membrane tension, and conclude that tension is not altering transition rates. Indeed, we previously found that membrane tension in whole cell mode did not significantly alter the phase or frequency response of electromotility (Santos-Sacchi and Song, 2014b). This is in contrast to salicylate which likely alters transition rates since NLC frequency response is affected (Santos-Sacchi and Tan, 2019).

Lastly, we revisited the data of Gale and Ashmore (Gale and Ashmore, 1997a), and recast our current and previous data with their prior results. The small variability within our NLC data set allowed us to identify a power dependence of frequency, whereas Gale and Ashmore originally applied a Lorentzian fit with a cut-off of 9.95 kHz. **Figure 5** shows that our data and power function fits reasonably describe the roll-off in their data, and highlights an inability of voltage to sufficiently drive prestin electromechanical activity at very high frequencies (60-160 kHz), where CA is known or expected to exist. This point is further emphasized by experiments showing that perilymphatic perfusion of 5 mM salicylate or its congener is devastating to CA (Santos-Sacchi et al., 2006; Fisher et al., 2012). Considering the 1.6 mM $K_{1/2}$ of salicylate action, the effect of the perilymphatic treatment is not a full block of NLC, but rather about 75% (12 dB) reduction (Kakehata and Santos-Sacchi, 1996). Thus, the reduction in NLC resulting from its power frequency dependence (12 dB down at 25 kHz, 20 dB down at 40 kHz, and 40 dB down at 53 kHz; **Figure 5**) has biological impact *across the full audio frequency spectrum of mammals*, and indicates a continuously diminishing influence of voltage alone to drive prestin electromechanical activity and CA via a cycle-by-cycle means. The additional low pass influence of external loads on the whole cell electromotility frequency response that we observed recently (Santos-Sacchi et al., 2019b), further lessens the effectiveness of voltage-dependent electromotility in influencing organ of Corti motion at very high frequencies. How CA is driven by OHCs in the basal high frequency turn of mammals that enjoy perception of sound above 60 kHz is an enigma at present.

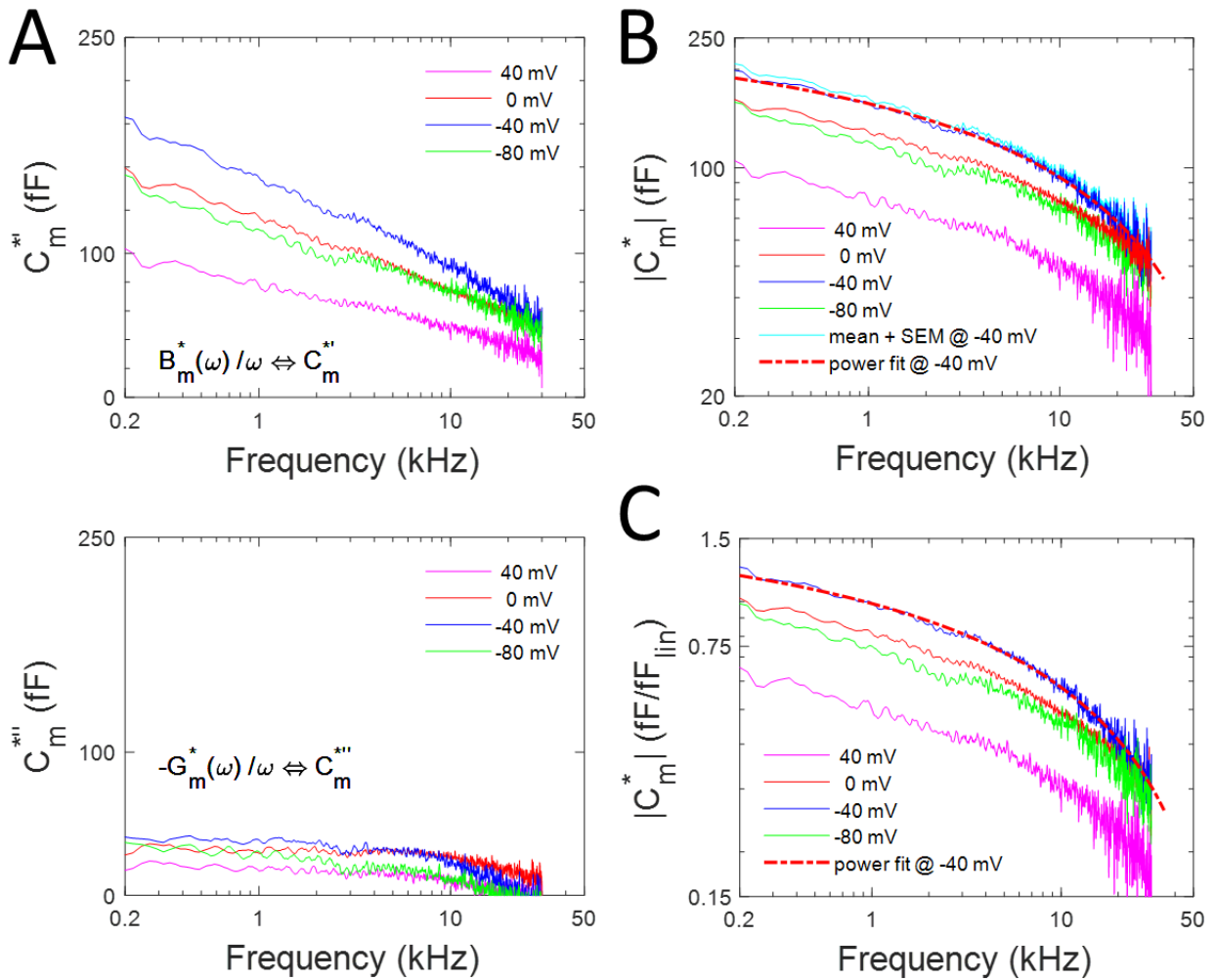

**Figure 1**. Complex capacitance of OHC lateral membrane macro-patches. **A)** Top panel shows the mean real components of complex capacitance at selected holding potentials; bottom panel shows corresponding imaginary parts. **B)** Plot of the mean magnitudes of complex capacitance at different holding potentials. A power function fit of the magnitude at the -40 mV holding potential is shown by the dotted line (red) overlying the mean (dark blue line). The light blue line is the mean + SEM at that same holding potential data. Fit parameters: $NLC_0$: 279.17 fF, a: -34.35, b: 0.18. **C)** Plot of the means of complex capacitance magnitude per estimated linear patch capacitance (see Methods). Fit parameters: $NLC_0$: 1.71 fF, a: -0.19, b: 0.19. Note low-pass behavior of magnitude functions.

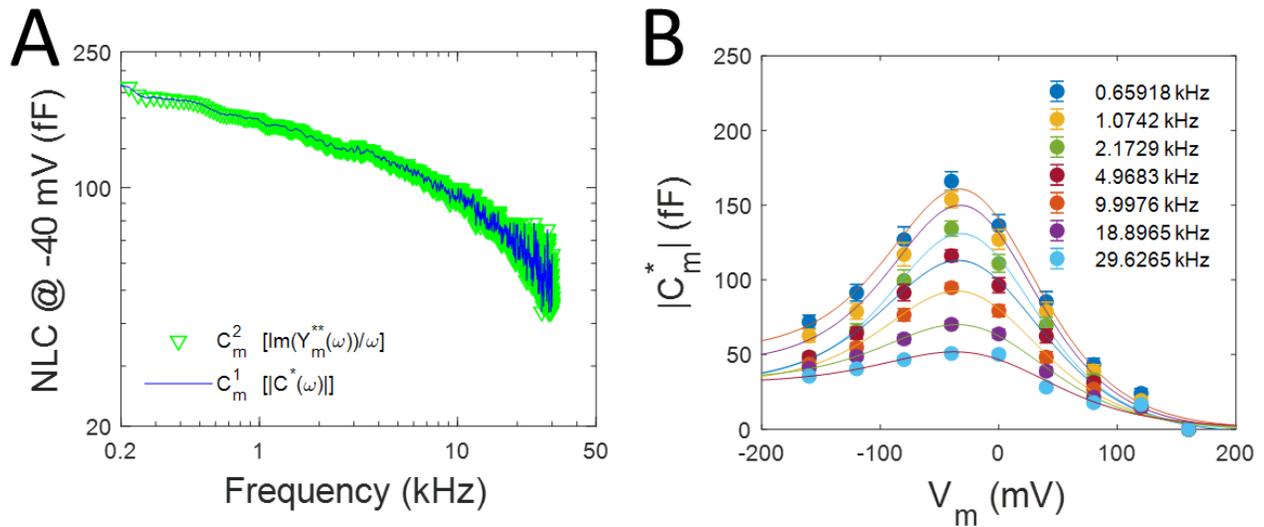

**Figure 2**. Comparison of two estimates of NLC frequency response at -40 mV holding potential. A) Each estimate provides equivalent results. B) C-V plot of complex capacitance magnitude of OHC lateral membrane macro-patches at selected frequencies (mean +/- SEM). Data are fit (*eq. m1*) to extract Boltzmann parameters (see Methods). -40 mV is close to $V_h$. From lowest to highest frequency fit values are (peak NLC, $V_h$, z): 133.60, 127.61, 115.42, 98.81, 76.81, 52.60, 34.35 fF; -25.2, -25.3, -26.0, -26.4, -26.9, -20.6, -16.4 mV; 0.61, 0.62, 0.62, 0.57, 0.59, 0.56, 0.54. Prior to averaging patch responses (n=25), capacitance values from each patch were obtained by averaging over a bandwidth of about 200 Hz about the listed frequency value.

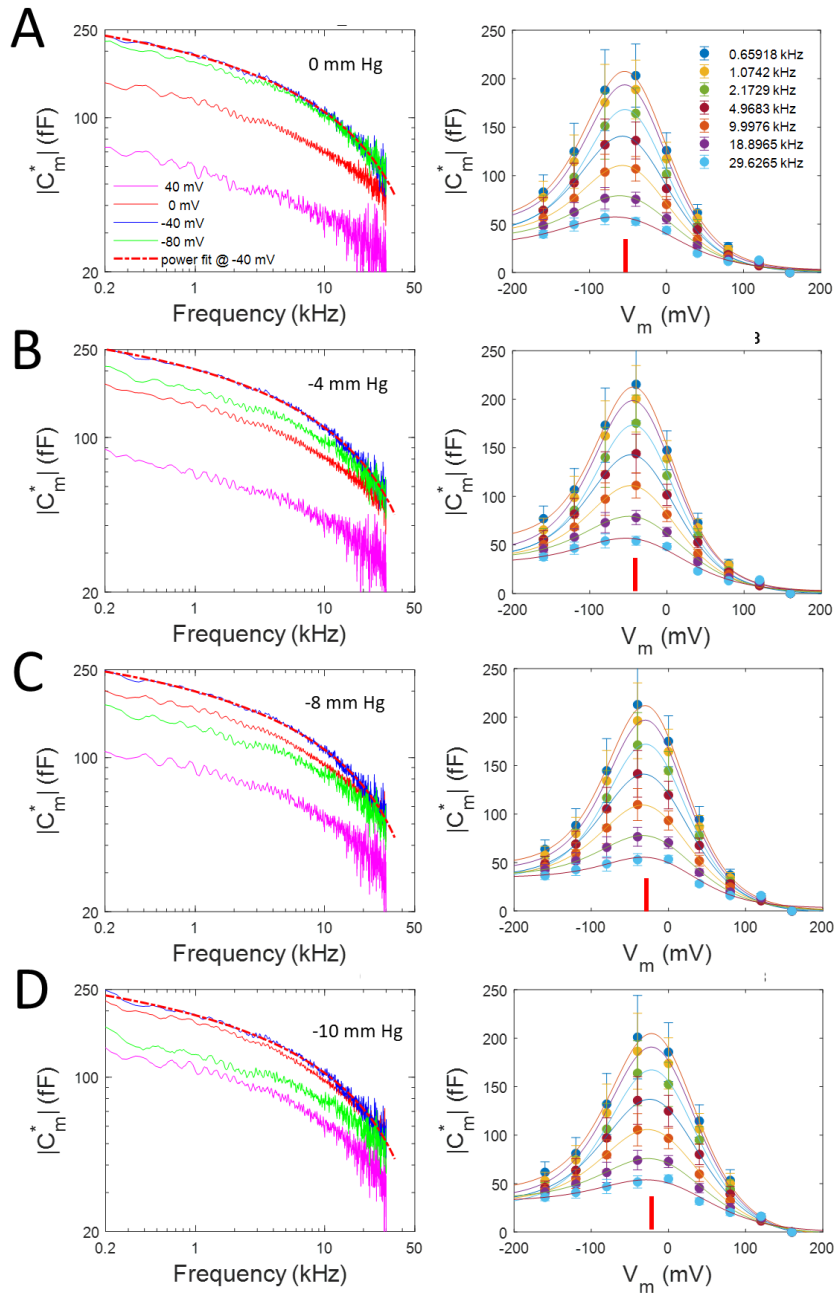

**Figure3**. Effect of membrane tension on magnitude of complex capacitance voltage and frequency dependence. **Panels A**, **B**, **C** and **D** depict results at 0, -4, -8 and -10 mm Hg pipette pressure. The effect of membrane tension is to shift $V_h$, as indicated by the change in voltage at maximal capacitance as negative pressure increases, without appreciably altering frequency response (dotted lines are fits to a power function). On the right, C-V plots of complex capacitance magnitude at the four different pipette pressures are shown. Red bars indicate voltage at peak capacitance. The depolarizing shift in $V_h$ is readily apparent as pressure increases. Peak magnitudes do not appreciably alter. $\Delta C_{sa}$ is tension and frequency independent. Mean +/- SEM (n=8). From lowest to highest frequency fit values from *eq. m1* are (peak NLC, $V_h$, z): **Panel A**, 179.15, 166.45, 146.43, 120.30, 88.90, 62.62, 41.51 fF; -49.4, -49.0, -49.0, -50.9, -49.0, -48.2, -51.2 mV; 0.70, 0.71, 0.71, 0.66, 0.70, 0.58, 0.53; **Panel B**, 182.80, 175.62, 153.91, 124.40, 91.56, 60.38, 39.08 fF; -38.5, -40.0, -39.4, -40.7, -39.4, -36.4, 35.8 mV; 0.74, 0.72, 0.72, 0.67, 0.66, 0.60, 0.57. **Panel C**, 187.73, 175.16, 155.35, 124.08, 90.68, 57.97, 34.77 fF; -25.9, -25.7, -25.7, -26.2, -26.0, -20.6, -14.5 mV; 0.73, 0.73, 0.72, 0.67, 0.66, 0.62, 0.65. **Panel D**, 185.03, 175.67, 156.00, 124.51, 91.22, 58.39, 36.32, fF; -17.2, -17.7, -18.0, -18.1, -8.8, -12.3, -8.0 mV; 0.67, 0.66, 0.66, 0.61, 0.60, 0.58, 0.55.

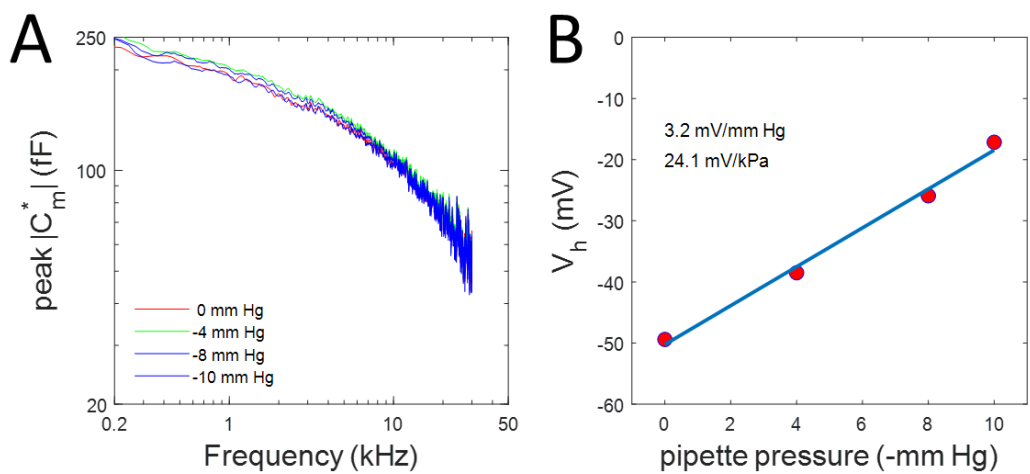

**Figure 4**. Effects of membrane tension on NLC frequency response and operating voltage. **A)** Comparison of NLC frequency response at -40 mV at four pipette pressures. Roll-off is equivalent as indicated by overlap of traces. **B)** $V_h$ of NLC at the lowest frequency in **Figure 3** is plotted versus pipette pressure. A linear fit (blue line) indicates a sensitivity of 24.1 mV/kPa.

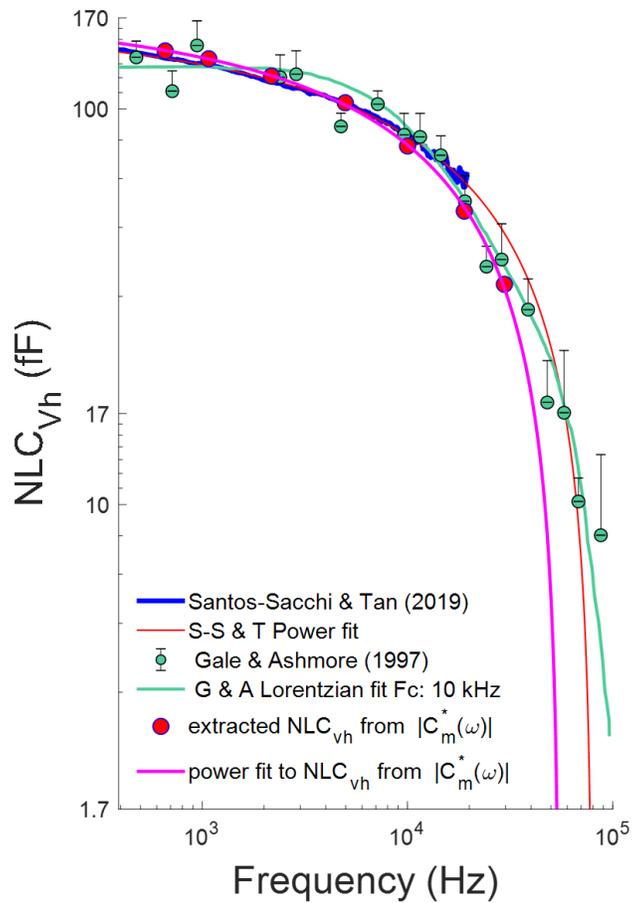

**Figure 5**. Replot of data from Santos-Sacchi and Tan (2019) with those of Gale and Ashmore (1997). Using a dual-sine capacitance estimation algorithm, NLC of OHC patches (blue line) exhibited a power dependence on frequency (red line). The data of Gale and Ashmore, collected with a traditional phase tracking approach using a lock-in amplifier (data: green circles; their fit green solid line) is commensurate with the dual-sine approach. Finally, NLC$_{vh}$ (red symbols) obtained from *eq. m1* fits of our complex capacitance data (from **Figure 2B**) are also commensurate with previous observations, giving a power fit (magenta line) with parameters NLC$_0$: 175.90 fF, a: -3.33, b: 0.36.

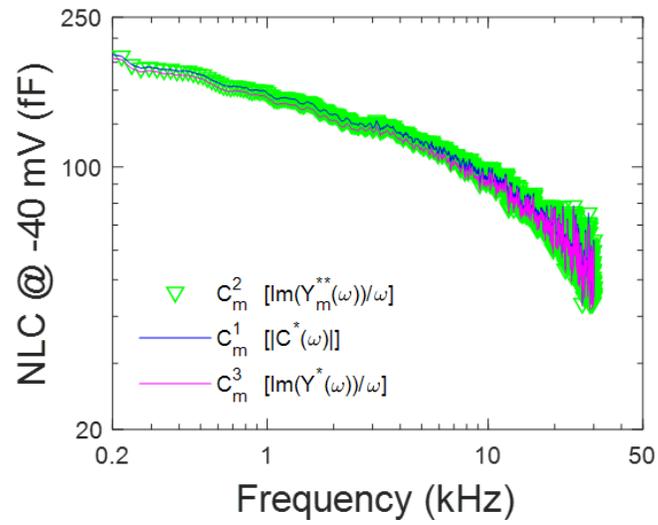

**Figure 6**. Three estimates of the frequency response of NLC at -40 mV holding potential. Estimates from raw admittance data (pink line) reasonably compare with the other estimates despite the fact that the real component is not taken into account.